\documentclass[12pt]{article}
\usepackage{amsmath}
\usepackage{graphicx}
\usepackage{color}
\usepackage{cite}
\usepackage{here}

\def \beq{\begin{equation}}
\def \eeq{\end{equation}}
\def\eqref#1{(\ref{#1})}
\def\bea{\begin{eqnarray}}
\def\eea{\end{eqnarray}}
\def\jpsi{J\kern-0.1em/\kern-0.1em\psi\kern0.03em}
\def\half{1\kern-0.15em/\kern-0.05em{2}}
\def\threehalves{3\kern-0.15em/\kern-0.05em{2}}
\def\nl{\hfill\break}

\def\URLtilde{\lower0.2em\hbox{$\tilde{\phantom{a}}$}}
\def\mycomm#1{\hfill\break\strut\kern-3em{\color{red}\tt ====> #1
\color{black}}\hfill\break}

\newcount\timecount
\newcount\hours \newcount\minutes  \newcount\temp \newcount\pmhours
\hours = \time
\divide\hours by 60
\temp = \hours
\multiply\temp by 60
\minutes = \time
\advance\minutes by -\temp
\def\hour{\the\hours}
\def\minute{\ifnum\minutes<10 0\the\minutes
\else\the\minutes\fi}
\def\clock{
\ifnum\hours=0 12:\minute\ AM
\else\ifnum\hours<12 \hour:\minute\ AM
\else\ifnum\hours=12 12:\minute\ PM
\else\ifnum\hours>12
\pmhours=\hours
\advance\pmhours by -12
\the\pmhours:\minute\ PM
\fi
\fi
\fi
\fi
}

\def\monthname{\relax\ifcase\month 0/\or January\or February\or
March\or April\or May\or June\or July\or August\or September\or
October\or November\or December\else\number\month/\fi}

\def\bold#1{\setbox0=\hbox{$#1$}     \kern-.025em\copy0\kern-\wd0
\kern.05em\copy0\kern-\wd0
\kern-.025em\raise.0433em\box0 }

\begin{document}
\setcounter{footnote}{1}
\rightline{EFI 22-7}
\vskip1.0cm

\centerline{\large \bf New strange pentaquarks}
\bigskip

\centerline{Marek Karliner$^a$\footnote{{\tt marek@tauex.tau.ac.il}}
 and Jonathan L. Rosner$^b$\footnote{{\tt rosner@hep.uchicago.edu}}}
\medskip

\centerline{$^a${\it School of Physics and Astronomy}}
\centerline{\it Raymond and Beverly Sackler Faculty of Exact Sciences}
\centerline{\it Tel Aviv University, Tel Aviv 69978, Israel}
\medskip

\centerline{$^b${\it Enrico Fermi Institute and Department of Physics}}
\centerline{\it University of Chicago, 5620 S. Ellis Avenue, Chicago, IL
60637, USA}
\bigskip
\strut

\begin{quote}
\begin{center}
ABSTRACT
\end{center}
The new strange pentaquarks observed by LHCb are very likely hadronic
mo\-le\-cules consisting of \,$\Xi_c\, \bar D$\,  
and \,$\Xi_c \,\bar D^{*}$.
We discuss the experimental evidence supporting this conclusion, pointing
out the similarities and differences with 
the $P_c(4312)$, $P_c(4440)$ and $P_c(4457)$ 
pentaquarks in the non-strange sector. 
The latter clearly are hadronic molecules consisting of 
\,$\Sigma_c\, \bar D$\, and \,$\Sigma_c \,\bar D^{*}$. 
Following this line of thought, we predict
three additional strange pentaquarks, consisting of 
\,$\Xi_c^{\prime}\, \bar D$\,  and \,$\Xi_c^{\prime} \,\bar D^{*}$.
The masses of these states are expected to be shifted upwards by 
\,$M(\Xi_c^{\prime})-M(\Xi_c) \approx 110$ MeV with
respect to the corresponding known strange pentaquarks.
\end{quote}
\smallskip

\bigskip


Very recently the LHCb Collaboration announced observation of a new
strange pentaquark 
$P^{\Lambda}_{\psi s}(4338)$\footnote{We employ here a new naming scheme
suggested by LHCb. An alternative name for this state is
$P_{cs}(4338)$.}
with minimal quark content $c \bar c u d s$,
mass $M = 4338.2 \pm 0.7$ MeV and width $\Gamma = 7.0 \pm 1.2$ MeV.
This new state has been observed 
in the decay
$B^- \to \jpsi \Lambda \bar p$
as a resonance in the $\jpsi \Lambda$ invariant mass 
with statistical significance
$> 10~\sigma$. 
Amplitude analysis yields spin-parity $J^P=\half^-$ with 
the alternative $J^P=\half^+$ rejected @90\% confidence level 
\cite{Penta4338}.
\bigskip

Several features of the new state are strongly suggestive \cite{Karliner:2015ina}
of a \,$\Xi_c \bar D$\, hadronic molecule:
\begin{itemize}
\item[(a)] Vicinity to the relevant baryon-meson threshold.
The central value of $P^{\Lambda}_{\psi s}(4338)$ mass is only
0.8 MeV above \,$\Xi_c^+ D^-$\, threshold and 2.9 MeV above 
\,$\Xi_c^0 D^0$\, threshold (cf. Appendix A).
\item[(b)] Spin and parity. The spin and parity of an 
$S$-wave hadronic molecule are necessarily inherited from its constituents.
In this case the latter are a positive parity spin-$\half$ baryon and a negative
parity spin-0 meson. $J^P=\half^-$ is exactly what is expected.
\item[(c)] Narrow width, compared with the phase space available for decay.
$P^{\Lambda}_{\psi s}(4338)$ decays into $\jpsi \Lambda$, whose threshold
is 4212.6 MeV, so the $Q$-value is 126 MeV. 
The 7 MeV width of $P^{\Lambda}_{\psi s}(4338)$ is unnaturally small for
such a $Q$-value, so there must be a suitable decay-suppressing mechanism at
work. Decay into $\jpsi \Lambda$ requires the charmed and anti-charmed
quarks getting close to each other, but in a $\Xi_c \bar D$ molecular
configuration the average distance between $\Xi_c$ and $\bar D$
is much larger than 1 fermi, automatically providing an efficient
decay-suppressing mechanism.
\end{itemize}

Additional (although less statistically significant) support 
for the molecular interpretation is provided by earlier
LHCb data on the $P^{\Lambda}_{\psi s}(4459)$ pentaquark
\cite{LHCb:2020jpq}, \cite{Karliner:2021xnq}.
In that case LHCb observed a strange pentaquark as a peak in
$\jpsi \Lambda$ invariant mass in the decay
$\Xi_b^- \to \jpsi \Lambda K^-$, with 
mass $M=4458.8 \pm 2.9^{+4.7}_{-1.1}$ MeV,
width $\Gamma = 17.3 \pm 6.5^{+8.0}_{-5.7}$ MeV
and statistical significance of 3.1 $\sigma$. 
The central value of the $P^{\Lambda}_{\psi s}(4459)$ mass is 
approximately 20 MeV below the $\Xi_c \bar D^*$ threshold.

Remarkably, LHCb observed \cite{LHCb:2020jpq}
that this resonance can equally well be described 
by {\em a two peak structure}, with the two peaks split by 13 MeV:
\bea
P^{\Lambda}_{\psi s}(4455): \qquad
M = 4454.9 \pm 2.7 \hbox{\ MeV},\quad \Gamma = 7.5 \pm 9.7 \hbox{\ MeV}
\nonumber
\\
\\
P^{\Lambda}_{\psi s}(4468): \qquad
M = 4467.8 \pm 3.7 \hbox{\ MeV},\quad \Gamma = 5.2 \pm 5.3 \hbox{\ MeV}.
\nonumber
\label{M1M2}
\eea
This pattern is consistent with general expectations (see, e.g., Refs.~%
\cite{Wu:2010jy,Chen:2016ryt,Shen:2019evi,Wang:2019nvm}).
For a recent review and additional references, see Ref.~\cite{Dong:2021bvy}.

The above structure is highly reminiscent of the two-peak pentaquark structure
discovered by LHCb \cite{LHCb:2019kea} in the non-strange sector,
following the original discovery of hidden-charm pentaquarks
\cite{LHCb:2015yax},
\bea
P^{N}_{\psi}(4440)^+: \qquad
  M = 4440.3 \pm 1.3^{+4.1}_{ -4.7} \hbox{\ MeV}, \quad   
\Gamma =20.6 \pm 4.9^{+8.7}_{-10.1} \hbox{\ MeV}
 \nonumber 
\\ 
\\
 P^{N}_{\psi}(4457)^+: \qquad
  M = 4457.3 \pm 0.6^{+4.1}_{-1.7} \hbox{\ MeV}, \quad   
\Gamma = \phantom{2}6.4 \pm 2.0^{+5.7}_{-1.9} \hbox{\ MeV}.
 \nonumber
\\ \nonumber
\\
(\hbox{a.k.a.\ } P_c(4440)^+ \hbox{ and\ } P_c(4457)^+) 
\phantom{aaaaaaaaaaaaaaaaaaaaaaaaaaaaaaa}
\nonumber
\label{non-strange-double-peak} \eea 
These two resonances are most likely the two possible spin states of
an $S$-wave hadronic molecule consisting of a spin-$\half$ \,$\Sigma_c$ and
spin-1 \,$\bar D^{*0}$.  Clearly, in that case the expected $J^P$ values
are \,$\half^-$ \,and\, $\threehalves^-$. 

Analogous reasoning leads to the expectation that the 
spin and parity of
$P^{\Lambda}_{\psi s}(4455)$ and $P^{\Lambda}_{\psi s}(4468)$
are the two possible values for  
an $S$-wave hadronic molecule consisting of a spin-$\half$ \,\,$\Xi_c$ and
spin-1 \,$\bar D^{*0}$, i.e., $\half^-$ and $3/2^-$.

In view of the above it is natural to interpret
$P^{\Lambda}_{\psi s}(4338)$
as the strange analogue of
$P^{N}_{\psi}(4312)^+$ 
also reported in \cite{LHCb:2019kea},
with $M=4311.9 \pm 0.7^{+6.8}_{-0.6}$ MeV and 
$\Gamma = 9.8 \pm 2.7^{+3.7}_{-4.5}$ MeV,
commonly interpreted as a $\Sigma_c \bar D$ hadronic molecule.

One remaining issue is the specific mechanism which provides attraction
between $\bar D$ and $\Xi_c$. Binding between $\bar D^*$ and $\Sigma_c$ or $\Xi_c$
can be provided by one-pion exchange. But since $\bar D$ is a pseudoscalar, 
its binding to another hadron cannot be provided by one-pion exchange, because 
that would require a vertex involving three pseudoscalars which is forbidden in QCD,
since such a vertex cannot simultaneously conserve parity and angular momentum.

In the case of a $\Sigma_c \bar D$ hadronic molecule a two-pion exchange can provide
binding, because the intermediate $\Lambda_c \bar D^*$ state is relatively close in
mass to the initial state \cite{MK_Bled}. Two-pion exchange is expected to be
weaker than one-pion exchange and as a result $P^{N}_{\psi}(4312)^+$ might be a
virtual state, rather than a fully-fledged bound state.

For $\Xi_c \bar D$ two-pion exchange is unlikely to work, since in this case the
intermediate state is too heavy. One relatively simple possibility is $\rho$-mediated
$t$-channel charge exchange,

\bea
\Xi_c^0 \, \bar D^0 \quad 
&\xrightarrow[\rho^-]{}&
\quad  \Xi_c^+ D^-
\nonumber\\
\\
\Xi_c^+ \bar D^- \quad 
&\xrightarrow[\rho^+]{}&
\quad  \Xi_c^0  \bar D^0
\nonumber
\label{rho_exchange}
\eea
The $\Xi_c \bar D$ state decays into $\Lambda\jpsi$, so it has isospin zero. 
In such a state $t$-channel $\rho$ exchange is attractive \cite{Wu:2007fc}.
Clearly, more quantitative statements require a specific model-dependent calculation.

At this point it is important to stress that the analogy 
between $\Sigma_c \bar D^{(*)}$ and $\Xi_c \bar D^{(*)}$ 
hadronic molecules goes only so far.
As discussed in Ref.~\cite{Karliner:2021xnq},
$P^{\Lambda}_{\psi s}(4455)$ and $P^{\Lambda}_{\psi s}(4468)$
{\em do not} correspond to an $SU(3)_F$ rotation $q \to s$ $(q=u,d)$ of
$P^{N}_{\psi}(4440)^+$ and $P^{N}_{\psi}(4457)^+$.
Neither does $P^{\Lambda}_{\psi s}(4338)$ correspond to an $SU(3)_F$ 
rotation of $P^{N}_{\psi}(4312)^+$.

The point is that in the non-strange pentaquark hadronic molecules
the charmed baryon is $\Sigma_c$, in which the two light quarks
form a ``bad diquark" $(ud)$, with spin-1 and isospin-1. An $SU(3)_F$
rotation $q \to s$ then takes the $\Sigma_c$ baryon to $\Xi_c^\prime$, rather
than to $\Xi_c$. 
The latter is approximately 110 MeV lighter than $\Xi^\prime$,\footnote{%
$M(\Xi_c^{\prime +})-M(\Xi_c^+) = 110.5 \pm 0.4$ MeV and
$M(\Xi_c^{\prime 0})-M(\Xi_c^0) = 108.3 \pm 0.4$ MeV, cf. Appendix.}
because in $\Xi_c$ the light quarks form a spin-0 $[qs]$ ``good diquark" 
which is significantly lighter than the spin-1 $qs$ ``bad diquark"
in $\Xi^\prime_c$.

Moreover, $\Xi_c^\prime$ cannot decay via the strong interaction, because
$M(\Xi_c^{\prime})-M(\Xi_c) < m_\pi$. It can only decay radiatively,
$M(\Xi_c^{\prime})\to M(\Xi_c)\, \gamma$. 
Thus from the point of view of strong interactions $\Xi_c^\prime$ is as stable
as $\Xi_c$.

\begin{figure}[t]
\begin{center}
\includegraphics[width=0.99\textwidth] {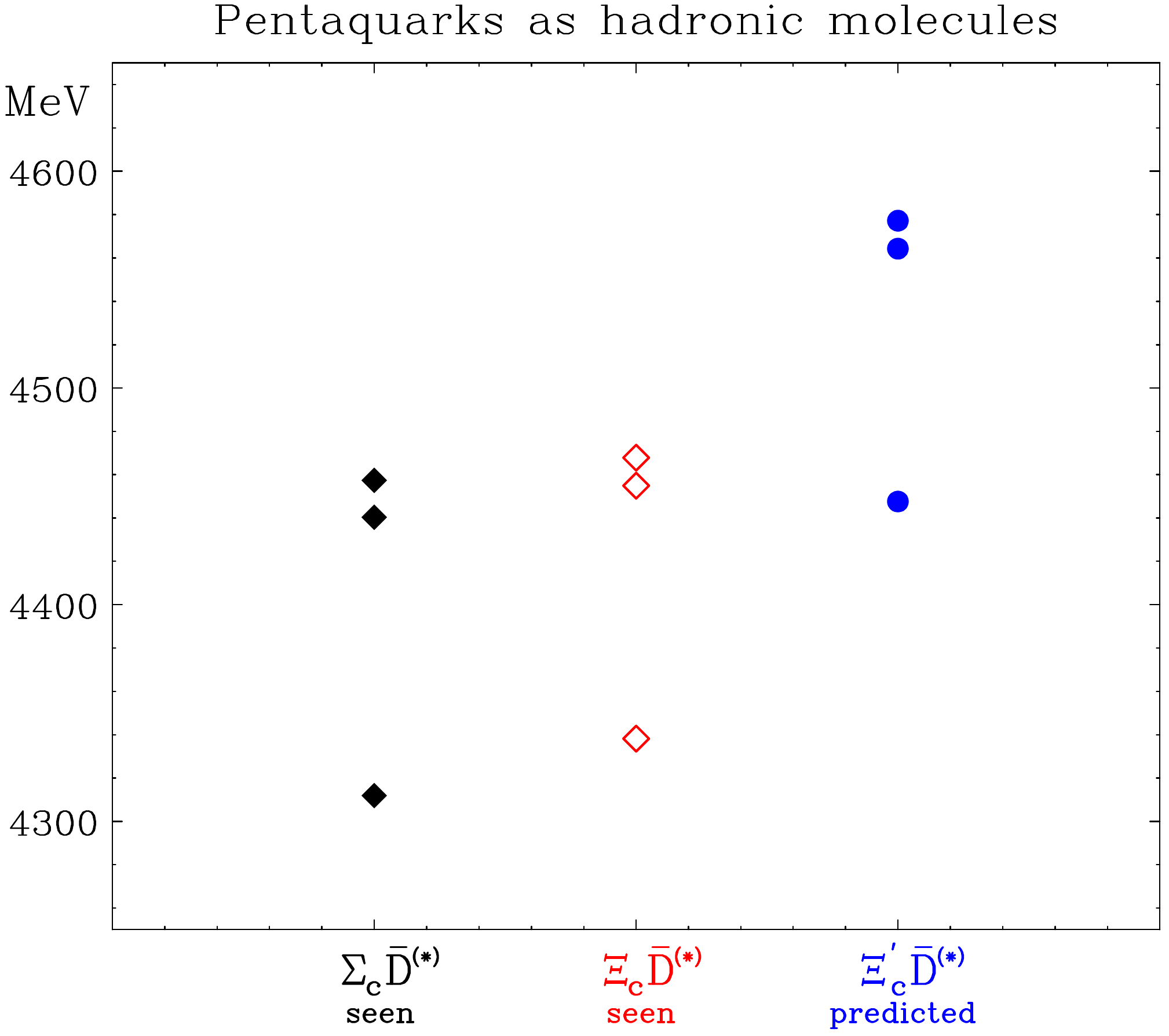}
\end{center}
\caption{Pentaquarks as hadronic molecules.
$\Sigma_c \bar D^{(*)}$ states are denoted by black diamonds,
$\Xi_c \bar D^{(*)}$ states by open red diamonds
and
$\Xi_c^\prime \bar D^{(*)}$ states by blue circles.
\label{fig:Pqs}}
\end{figure}

The upshot of the above observations is that, if --- as strongly hinted by the
data ---
$P^{\Lambda}_{\psi s}(4338)$
$P^{\Lambda}_{\psi s}(4455)$ and $P^{\Lambda}_{\psi s}(4468)$
indeed are 
$\Xi_c \bar D$ and $\Xi_c \bar D^{*}$ hadronic molecules,
then one should expect analogously three additional narrow strange pentaquarks,
corresponding to
$\Xi_c^\prime \bar D$ and $\Xi_c^\prime \bar D^{*}$ hadronic molecules.
Their masses are expected to be shifted by
$M(\Xi_c^{\prime})-M(\Xi_c) \approx 110$ MeV with
respect to the corresponding known strange pentaquarks, 
putting them approximately at 4448, 4564 and 4577 MeV,
as shown in Fig.~\ref{fig:Pqs}.
Their spin-parity quantum numbers are expected to be the same as those of their
counterparts.
Their widths are expected to be rather small, similar to those of
$P^{\Lambda}_{\psi s}(4338)$,
$P^{\Lambda}_{\psi s}(4455)$ and $P^{\Lambda}_{\psi s}(4468)$.

A potentially challenging point is that the 
$\Xi_c^\prime \bar D$ state at 4448 MeV, analogous to 
$P^{\Lambda}_{\psi s}(4338)$,
is expected just 7 MeV below $P^{\Lambda}_{\psi s}(4455)$.
This is because $\bar D^*- \bar D$ splitting plus the $\Xi_c \bar
D^*$ binding energy is close to $\Xi_c^\prime-\Xi_c$ splitting.
$\Xi_c^\prime \bar D$ state is expected to have spin-$\half$, so
if $P^{\Lambda}_{\psi s}(4455)$ turns out to also have spin-$\half$, the
two states will likely mix. 

\section*{Summary}
Recently
LHCb has reported several new narrow strange pentaquarks decaying into $\Lambda
\jpsi$, with minimal quark content $c \bar c u d s$. 
We have reviewed the experimental evidence and theoretical arguments 
strongly suggesting that they are \,$\Xi_c \,\bar D^{(*)}$\
hadronic mo\-le\-cules. The main points are their
proximity to the relevant baryon-meson thresholds, spin-parity and
unnaturally narrow widths, given the phase space available for decay.

We have discussed their similarities and differences with the three
nonstrange narrow pentaquarks decaying into $p \jpsi$, 
with minimal quark content $c \bar c u u d$, 
reported by LHCb in 2019. 

On the basis of this discussion, we predict three
additional narrow strange pentaquarks, corresponding to
\,$\Xi_c^\prime \,\bar D^{(*)}$\ hadronic molecules, with masses shifted
upwards by approximately 110 MeV with respect to the known
\,$\Xi_c \,\bar D^{(*)}$\ states, i.e.,
approximately at 4448, 4564 and 4557 MeV and with narrow widths.

\section*{ACKNOWLEDGMENTS}
The research of M.K. was supported in part by NSFC-ISF grant No.\ 3423/19.

\section*{APPENDIX}
\def\strutA{\vrule width 0pt height 2.5ex depth 1ex}
\begin{table}[H]
\begin{center}
\begin{tabular}{|c|c|}
\hline
state & mass (MeV) \cite{PDG2022} \\
\hline
\hline
$\Sigma_c^+$          &  $2452.65^{+0.22}_{-0.16}$                     \strutA \\ \hline
$\Sigma_c^0$          &  $2453.75^{+0.14}_{-0.14}$                     \strutA \\ \hline
\hline
$\Xi_c^+$             &  $2467.71^{+0.23}_{-0.23}$                     \strutA \\ \hline
$\Xi_c^0$             &  $2470.44^{+0.28}_{-0.28}$                     \strutA \\ \hline
\hline
$\Xi_c^{\prime +}$    &  $2578.2\,\phantom{4}^{+0.5\,\,}_{-0.5\,\,}$   \strutA \\ \hline
$\Xi_c^{\prime 0}$    &  $2578.7\,\phantom{4}^{+0.5\,\,}_{-0.5\,\,}$   \strutA \\ \hline
\hline
$\bar D^0$            &  $1864.84\,\,^{+0.05}_{-0.05}$   \strutA \\ \hline
$D^-$                 &  $1869.66\,\,^{+0.05}_{-0.05}$   \strutA \\ \hline
\hline
$\bar D^{*0}$         &  $2006.85\,\,^{+0.05}_{-0.05}$   \strutA \\ \hline
$D^{*-}$              &  $2010.26\,\,^{+0.05}_{-0.05}$   \strutA \\ \hline
\end{tabular}
\end{center}
\caption{Masses of charmed hadrons discussed in the text.
\label{appendix_masses}}
\end{table}

\end{document}